\RequirePackage{silence}
\documentclass[aps,prb,reprint]{revtex4-1}
\usepackage{amsmath}  
\usepackage{amsfonts} 
\usepackage{amssymb}
\usepackage{graphicx} 
\usepackage[american]{babel}
\usepackage[utf8]{inputenc}
\usepackage{natbib}
\usepackage{hyperref}

\renewcommand{\vec}[1]{\boldsymbol{\mathbf{#1}}} 
\newcommand{\m}[2]{m_{{#1}\vert{#2}}}
\newcommand{\Gal}{\mathtt{Gal}}
\newcommand{\IGal}{\mathtt{IGal}}
\newcommand{\reals}{\mathbb{R}}
\newcommand{\SO}{\mathrm{SO(3)}}
\newcommand{\GL}{\mathrm{GL}}
\newcommand{\ST}{\mathcal{M}}
\newcommand{\Kin}{\mathcal{K}}
\newcommand{\Dyn}{\mathcal{D}}

\begin{document}

\title{What if active and passive gravitational 
masses were not equal?}

\author{Domenico Giulini}
\affiliation{Institute for 
Theoretical Physics\\
Leibniz University of Hannover\\
Appelstraße 2, 30167 Hannover, Germany}

\date{\today}
\begin{abstract}
At first glance, combining Newton’s laws of 
motion with his law of gravitation 
seems straightforward. Students learn to 
distinguish inertial from gravitational 
mass and that their empirical equality is 
a remarkable fact about nature that 
will later serve as the conceptual 
gateway to general relativity.
However, a closer look reveals a further 
and often neglected distinction within 
Newtonian gravity: that between 
active and passive gravitational mass. 
A common textbook argument maintains that 
these must be equal, for otherwise Newton's 
third law would be violated. We review and 
critically re-examine this familiar 
reasoning and show that the supposed 
theoretical proof is not compelling. 
Our analysis highlights subtle structural
assumptions within Newtonian mechanics 
and offers physics teachers and researchers
a fresh opportunity to explore foundational 
questions with potentially interesting
applications in observational astronomy. 
In addition, an extended appendix, which is 
not part of the published AJP paper, offers 
some mathematical background material 
on the Galilean group and its action as 
group of dynamical symmetries for the type of dynamical equations considered here.   
\end{abstract}

\maketitle

\section{Introduction}
Presumably Hermann Bondi was the first 
person to point out in 1957 that Newton’s 
laws of motion combined with Newton's theory 
of gravitation gives rise to three concepts 
of mass that should be carefully
distinguished.\cite{Bondi:1957}
Let us start by briefly recalling what 
they are. 

Newton's second law states that any 
force $\vec F$ acting on a (pointlike) 
body leads to a change in its state of 
motion. It also states that this change 
is characterized by an acceleration 
$\vec a=\ddot{\vec x}$ proportional to 
$\vec F$. The constant of proportionality 
is called the body's \emph{inertial mass} 
$m_i$; hence
\begin{equation}
\label{eq:NewtonSecond}
\vec F=m_i\,\vec a=m_i\,\ddot{\vec x}\,.
\end{equation}
    
Newton's law of gravitation states that 
a (pointlike) body placed at position
$\vec y$ in space produces a 
gravitational field $\vec g$ whose 
strength at position $\vec x$ is 
\begin{equation}
\label{eq:NewtonGravity-a}
\vec g(\vec x)=G\,m_a\,\frac{\vec y-\vec x}{\ \vert\vec y-\vec x\vert^3}\,.
\end{equation}  
Here $G$ is Newton's universal constant and 
$m_a$ is the body's \emph{active gravitational 
mass}. This name is chosen because $m_a$ 
characterizes the body's ability to act as 
the source of a gravitational field. 

Finally, given a gravitational field 
$\vec g$ and a (pointlike) test-body 
at position $\vec x$. The force felt 
by the body will be proportional to 
the gravitational field at the position 
of the body, 
\begin{equation}
\label{eq:NewtonGravity-p}
\vec F=m_p\,\vec g(\vec x)\,.
\end{equation}  
The constant of proportionality, $m_p$, 
is called the body's \emph{passive 
gravitational mass}. 

From observations of suitable quantities 
one may then hope to obtain upper bounds 
on the possible inequality of the masses 
attributed to a body. Assuming for the 
moment a strict equality between active 
and passive gravitational mass, i.e.
$m_a=m_p=:m_g$, any variation of the 
ratio $m_i/m_g$ between physical bodies 
would lead to different free-fall 
orbits with identical initial 
conditions and hence to a violation of
the universality of free fall (UFF), 
also known as the weak equivalence 
principle (WEP). WEP lies at the 
foundation of General Relativity and 
has been tested again and again. 
Upper bounds on relative variations 
are currently at the $10^{-15}$ level, 
like in the recently completed 
MICROSCOPE mission.\cite{Touboul.EtAl:2022} 

In contrast, far fewer observations 
exist concerning the equality of 
$m_a$ and $m_p$, though meanwhile their 
accuracy almost matches that of 
WEP.\cite{Kreuzer:1968,Bartlett:1986,Singh.EtAl:2023}
The considerably smaller interest devoted 
in the literature to the question of the 
equality of the two gravitational masses 
is almost certainly connected with the 
widespread belief that equality allegedly 
follows from the equations themselves for 
reasons of internal consistency. The present 
paper examines this claim in some detail.  

We start with a review of the argument in 
section\,\ref{sec:GTBP1}. In section\,\ref{sec:GTBP2} 
we show that the gravitational two (or $N$) 
body problem is, in fact, completely free 
of any of the anomalies mentioned, so that 
no issue of inner consistency exists. 
In sections\,\ref{sec:SGH1} and \ref{sec:SGH2} 
we consider compound bodies held in rigid 
shape by non gravitational forces. In that 
case anomalies may exist, depending on the 
nature of the binding forces. All this leads 
to a deeper understanding of Newtonian concepts 
on the level of an introductory course on 
analytical mechanics.

\section{The gravitational two-body 
problem: first formulation
\label{sec:GTBP1}
}
Consider two (pointlike) bodies, 
labelled by 1 and 2, at positions 
$\vec x_{1{,}2}$ and inertial, active, 
and passive masses $\m{1{,}2}{i}$,
$\m{1{,}2}{a}$, and $\m{1{,}2}{p}$.
The gravitational force with which 
body\,1 is acted upon by body\,2 will be 
denoted by $\vec F_{12}$. Likewise, 
$\vec F_{21}$ denotes the force with 
which body\,2 is acted upon by body\,1. Combining \eqref{eq:NewtonSecond}, 
\eqref{eq:NewtonGravity-a}, and 
\eqref{eq:NewtonGravity-p},
we get the full set of equations of 
motion for the gravitational two-body problem: 
\begin{subequations}
\label{eq:NewGrav}
\begin{alignat}{2}
\label{eq:NewGrav-a}
\m{1}{i}\ddot{\vec x}_1
&\,=\,\vec F_{12}
&&\,=\,
G\,\m{1}{p}\m{2}{a}
\frac{{\vec x}_2-{\vec x}_1}{\vert{\vec x}_2-{\vec x}_1\vert^3}\,,\\
\label{eq:NewGrav-b}
\m{2}{i}\ddot{\vec x}_2
&\,=\,
\vec F_{21}
&&\,=\,
G\,\m{2}{p}\m{1}{a}
\frac{{\vec x}_1-{\vec x}_2}{\vert{\vec x}_1-{\vec x}_2\vert^3}\,,
\end{alignat}
\end{subequations}  
where, e.g., $\vert\vec x\vert$ denotes the Euclidean norm of $\vec x\in\reals^3$.  
Adding \eqref{eq:NewGrav-a} and \eqref{eq:NewGrav-b} we get for the terms 
on the left 
\begin{equation}
\label{eq:AddedLHS}
\m{1}{i}\ddot {\vec x}_1
+\m{2}{i}{\ddot{\vec x}}_2=
\frac{d\vec P}{dt}
\end{equation}
where 
\begin{equation}
\label{eq:TotalMomentum}
\vec P:=\vec p_1+\vec p_2
=\m{1}{i}\dot{\vec x}_1
+\m{2}{i}\dot{\vec x}_2
\end{equation}
is the system's total momentum. On the right-hand side we get 
\begin{equation}
\label{eq:SumGravForces}
{\vec F}_{12}+{\vec F}_{21}=
-S_{12}\,G\frac{\m{1}{p}\m{2}{p}}{\vert\vec x_1-\vec x_2\vert^2}\ \vec n\,,
\end{equation}
where 
\begin{equation}
\label{eq:NormalVector}
\vec n:=
\frac{\vec x_2-\vec x_1}%
{\vert\vec x_2-\vec x_1\vert}
\end{equation}
is the unit vector pointing from 
particle 1 to particle 2. Moreover, 
\begin{equation}
\label{eq:EoetvoesFactor}
S_{12}:=\frac{\m{1}{a}}{\m{1}{p}}
-\frac{\m{2}{a}}{\m{2}{p}}
\end{equation}
is a dimensionless parameter that 
measures the degree to which the 
quotient between active and passive 
gravitational mass is not universal.
\cite{Bartlett:1986,Singh.EtAl:2023}

Equality of \eqref{eq:AddedLHS} with 
\eqref{eq:SumGravForces} now implies 
that the total momentum changes in time 
if $S_{12}\ne 0$, i.e. if the ratio of 
the active and passive gravitational 
mass is not a universal constant. 
Clearly, in our context, the observed 
momentum non-conservation directly 
results from a violation of Newton's 
third law (action equals reaction) 
which would require the sum of forces
 \eqref{eq:SumGravForces} to vanish. 

Another way to express the same state 
of affairs is to use the 
center-of-inertial-mass: 
\begin{equation}
\label{eq:CoM-inertial}
\vec X_i:=\frac{\m{1}{i}\vec x_1+\m{2}{i}\vec x_2}%
{\m{1}{i}+\m{2}{i}}\,.
\end{equation}
Equality between \eqref{eq:AddedLHS} and 
\eqref{eq:SumGravForces} then reads 
\begin{equation}
\label{eq:AccInertialCoM}
\ddot{\vec X}_i=
S_{12}\, G\,\frac{\m{1}{p}\m{2}{p}}{\m{1}{i}+\m{2}{i}}
\,\frac{\vec x_1-\vec x_2}{\vert\vec x_1-\vec x_2\vert^3}\,,
\end{equation}
showing that the (inertial!) center-of-mass 
suffers a non-vanishing acceleration if 
$S_{12}\ne 0$.

All this seems to suggest that the equations 
of motion \eqref{eq:NewGrav} violate 
basic theoretical principles of Newtonian 
dynamics, like Newton's third law, 
unless active and passive gravitational 
masses were universally equal.\cite{Laemmerzahl.EtAt:2007,Erling:2023}.

This impression is further reinforced by 
the simple observation that the set of 
equations 
\eqref{eq:NewGrav} are clearly 
invariant under spatial translations.
This means that if ${\vec x}_1(t)$ and 
${\vec x}_2(t)$ solve \eqref{eq:NewGrav}, 
so do ${\vec x'}_1(t):={\vec x}_1(t)+\vec a$
and ${\vec x'}_2(t):={\vec x}_2(t)+\vec a$
for any time-independent vector $\vec a$.
(In fact, the whole 10-parameter Galilean 
group acts as symmetries for these equations
-- compare the Mathematical Appendix--, 
though that fact is not needed here.) 
But, according to the Noether Theorem
\cite{Desloge.Karch:1977,Neuenschwander:2014,Scheck:Mechanics},
space-translation symmetry should give 
rise to momentum conservation if the 
equations of motion derive from an invariant 
variational principle.\footnote{The 
variational principle is also known as 
the \emph{principle of stationary action}. 
Transformations acting on the dynamical 
trajectories leave the action invariant 
if  the Lagrangian changes at most by a 
total time derivative. This also implies 
the invariance of the Euler-Lagrange 
equations of motion; see 
Ref.\citenum{Desloge.Karch:1977} for a 
straightforward proof. Hence solution
curves are mapped to solution curves
of the very same set of equations.
However, the converse need not be 
true: symmetries of the equations 
of motion need not leave the Lagrangian 
action invariant up to a total time derivative. 
Those that do are called \emph{Noether symmetries}}. 
So are we to conclude that the system
\eqref{eq:NewGrav} of equations is not 
of Euler-Langrange form (has no variational 
formulation)? 

As we will see in the next section, the 
difficulties just mentioned do not lie 
in the equations themselves, but rather in 
a prejudiced expectation concerning the 
analytic expression for dynamic quantities,
like momentum. We mention in passing that,
in other contexts, controversial discussions 
on alleged mismatches between derived analytic 
expressions for conserved quantities and their 
expected form are not uncommon in the literature.\cite{Lemos:2021,McDonald:2022,Lemos:2022}

\section{The gravitational two-body 
problem: second formulation \label{sec:GTBP2}
} 
We rewrite the system \eqref{eq:NewGrav} 
by multiplying \eqref{eq:NewGrav-a}
by $\m{1}{a}/\m{1}{p}$ and 
\eqref{eq:NewGrav-b} by $\m{2}{a}/\m{2}{p}$, 
leading to 
\begin{subequations}
\label{eq:NewGravAlt}
\begin{alignat}{1}
\label{eq:NewGravAlt-a}
\mu_1\ddot{\vec x}_1
&\,=\,G\,\m{1}{a}\m{2}{a}
\frac{{\vec x}_2-{\vec x}_1}{\vert{\vec x}_2-{\vec x}_1\vert^3}\,,\\
\label{eq:NewGravAlt-b}
\mu_2\ddot{\vec x}_2
&\,=\,G\,\m{1}{a}\m{2}{a}
\frac{{\vec x}_1-{\vec x}_2}{\vert{\vec x}_1-{\vec x}_2\vert^3}\,,
\end{alignat}
\end{subequations}  
where 
\begin{equation}
\label{eq:DefMuMasses}
\mu_1:=
\m{1}{i}\frac{\m{1}{a}}{\m{1}{p}}\quad\text{and}\quad
\mu_2:=
\m{2}{i}\frac{\m{2}{a}}{\m{2}{p}}\,.
\end{equation}
The right-hand sides of 
\eqref{eq:NewGravAlt} are now equal 
in magnitude and opposite in direction, 
so that their sum yields  
\begin{equation}
\label{eq:MuMomentumCons-1}
\mu_1\ddot{\vec x}_1+\mu_2\ddot{\vec x}_2=\vec 0\,.
\end{equation}
This is equivalent to the statement 
that 
\begin{equation}
\label{eq:MuCoM}
{\vec X}_\mu:=\frac{\mu_1\vec x_1+\mu_2\vec x_2}{\mu_1+\mu_2}
\end{equation}
is free of acceleration, i.e. moves in a straight 
line. We call ${\vec X}_\mu$ the center-of-$\mu$-mass, 
which is to be distinguished from the center-of-inertial-mass, given by \eqref{eq:CoM-inertial}. Their difference 
is a multiple of the relative position 
\begin{equation}
\label{eq:DefRelPos}
\vec r:=\vec x_2-\vec x_1\,.
\end{equation}
Indeed, using \eqref{eq:MuCoM} and \eqref{eq:DefRelPos} 
we can solve for $\vec x_{1{,}2}$:
\begin{subequations}
\label{eq:PositionsExpressed}
\begin{alignat}{1}
\label{eq:PositionsExpressed-a}
\vec x_1
&:={\vec X}_\mu-\frac{\mu_2}{\mu_1+\mu_2}\,\vec r\,,\\
\label{eq:PositionsExpressed-b}
\vec x_2
&:={\vec X}_\mu+\frac{\mu_1}{\mu_1+\mu_2}\,\vec r\,.
\end{alignat}
\end{subequations} 
Inserting this into \eqref{eq:CoM-inertial}, using  
\eqref{eq:DefMuMasses} and the 
definition \eqref{eq:EoetvoesFactor} of 
$S_{12}$, we obtain
\begin{equation}
\label{eq:CoM-Relation}
\vec X_i={\Vec X}_\mu+S_{12}\frac{\m{\rm red}{i}}{\mu_{\rm tot}}
\,\vec r\,.
\end{equation}
Here and in the following we use the 
standard definitions for total and reduced masses:
\begin{subequations}
\label{eq:DefMassFunctions}
\begin{alignat}{2}
\label{eq:DefMassFunctions-a}
\m{\rm tot}{i}
&:=\m{1}{i}+\m{2}{i}\quad
&&\text{(total inertial mass)}\,,\\
\label{eq:DefMassFunctions-b}
\m{\rm red}{i}
&:=\frac{\m{1}{i}\m{2}{i}}{\m{1}{i}
    +\m{2}{i}}\quad
&&\text{(reduced inertial mass)}\,,\\
\label{eq:DefMassFunctions-c}
\mu_{\rm tot}
&:=\mu_1+\mu_2\quad
&&\text{(total $\mu$-mass)}\,,\\
\label{eq:DefMassFunctions-d}
\mu_{\rm red}
&:=\frac{\mu_1\mu_2}{\mu_1+\mu_2}
&&\text{(reduced $\mu$-mass)}\,.
\end{alignat}
\end{subequations}

We can summarize the content of this 
section in the now seemingly trivial 
observation that the differential 
equations of Newton’s two-body 
problem with general mass triples 
$(\m{1}{i},\m{1}{p},\m{1}{a})$ and 
$(\m{2}{i},\m{2}{p},\m{2}{a})$ are 
identical to those with inertial masses 
 $\mu_1$ and $\mu_2$ and gravitational 
masses $\m{1}{a}$ and $\m{2}{a}$, 
both active and passive! In particular it follows
that, in the gravitational 2-body 
problem, orbital data alone cannot 
distinguish these two sets of masses 
and are therefore insensitive to possible 
differences in active and passive 
gravitational masses. 

Having said this, it is now obvious that 
equations \eqref{eq:NewGravAlt} can be derived
from an action principle whose Lagrangian $L$ 
is just, as usual, of the form $L=T-V$, where $T$ 
is the kinetic and $V$ the potential energy. 
Hence 
\begin{equation}
\label{eq:LagrangeFunction}
L=\frac{1}{2}\bigl(
\mu_1\vert\dot{\vec x}_1\vert^2+\mu_2\vert\dot{\vec x}_2\vert^2\bigr)
+G\,\frac{\m{1}{a}\m{2}{a}}{\vert{\vec x}_1-\vec{x}_2\vert}\,.
\end{equation} 
Alternatively, using \eqref{eq:PositionsExpressed},
it can be written in terms of the variables 
$({\Vec X}_\mu,\vec r)$:
\begin{equation}
\label{eq:LagrangeFunction-2}
L=\frac{1}{2}\bigl(\mu_{\rm tot}\vert
{\dot{\vec X}}_\mu\vert^2+\mu_{\rm red}\vert
\dot{\vec r}\vert^2\bigr)
+G\,\frac{\m{1}{a}\m{2}{a}}{r}\,,
\end{equation}
where $r:=\vert\vec r\vert$. 
The Euler-Lagrange equations 
for \eqref{eq:LagrangeFunction-2} are 
\begin{subequations}
\label{eq:EL-Eq-Alt}
\begin{alignat}{1}
\label{eq:EL-Eq-Alt-a}
{\ddot{\vec X}}_\mu
&=\Vec 0\,,\\
\label{eq:EL-Eq-Alt-b}
\ddot{\vec r}
&=-G\frac{\m{1}{a}\m{2}{a}}{\mu_{\rm red}}\frac{\vec r}{r^3}\,.
\end{alignat}
\end{subequations}
Differentiating \eqref{eq:CoM-Relation} twice
and inserting \eqref{eq:EL-Eq-Alt} gives
\begin{subequations}
\label{eq:MotionInertialCoM}
\begin{equation}
\label{eq:MotionInertialCoM-a}
{\ddot{\vec X}}_i=-GM\,S_{12}\frac{\vec r}{r^3}\,, 
\end{equation}
with
\begin{equation}
\label{eq:MotionInertialCoM-b}
M:=\frac{\m{\rm red}{i}\m{1}{a}\m{2}{a}}{\mu_{\rm red}\mu_{\rm tot}}
= \frac{\m{1}{p}\m{2}{p}}{\m{\rm tot}{i}}\,.
\end{equation}
\end{subequations}
Clearly, \eqref{eq:MotionInertialCoM-a} is just \eqref{eq:AccInertialCoM}.

$L$ is obviously invariant under space-time 
translations and rotations, whereas 
under boost-transformation it only 
changes by a total time derivative.
\footnote{This is because under boost
transformations $\vec x\mapsto\vec x+\vec vt$   
squared velocities, which appear in the 
kinetic-energy terms, change like 
$\dot{\vec x}^2\mapsto
\dot{\vec x}^2+d/dt(2\vec x\cdot\vec v+t\vec v^2)$.} 
Hence the whole 10-parameter\footnote{There is one 
parameter for time translations and 
three parameters each for space translations, 
boosts, and rotations. The four-parameter set of 
translations are called the ``inhomogeneous'' 
transformations.} set of inhomogeneous Galilean 
transformations acts by Noether symmetries. 
At this point we refer to the Mathematical Appendix 
for more background information on the Galilean 
group and what it means to say that it acts by 
dynamical symmetries on the solutions of a given 
set of differential equations for $N$ point 
particles.
  
As a consequence of the Noether symmetries there 
exist 10 conserved quantities, the systematic 
derivation of which may, e.g., 
be found in Ref.\citenum{Desloge.Karch:1977}
or Sec.\,2.41 of Ref.\citenum{Scheck:Mechanics}.
They correspond to the overall energy $E$ 
(time translations), the overall momentum 
$\vec P$ (space-translations), the overall 
angular momentum $\vec J$ (spatial rotations), 
and the center-of-mass $\vec C$ at $t=0$  
(boosts). Their analytic expressions are as 
follows:
\begin{subequations}
\label{eq:ConsQuant}
\begin{alignat}{1}
\label{eq:ConsQuant-a}
E
&=\frac{1}{2}\bigl(
\mu_{\rm tot}\vert{\dot{\vec X}}_\mu\vert^2+\mu_{\rm red}\vert\dot{\vec r}\vert^2\bigr)
-G\,\frac{\m{1}{a}\m{2}{a}}{r}\,,
\\
\label{eq:ConsQuant-b}
\vec P
&=\mu_{\rm tot}{\dot{\vec X}}_\mu\,,\\
\label{eq:ConsQuant-c}
\vec J
&=\mu_{\rm tot}{\vec X}_\mu\times
{\dot{\vec X}}_\mu+\mu_{\rm red}\,\vec r\times\dot{\vec r}=:\vec L+\vec S\,,\\
\label{eq:ConsQuant-d}
\vec C
&={\vec X}_\mu-\frac{\vec P}{\mu_{\rm tot}}t\,. 
\end{alignat}
\end{subequations}
Note that the total angular momentum 
$\vec J$ splits as sum of its orbital 
part $\vec L$ and its spin part $\vec S$,
each of which is conserved individually. 
This immediately follows from
${\ddot{\vec X}}_\mu=\vec 0$ and 
$\ddot{\vec r}\propto\vec r$. 

In summary, we can say that the gravitational 
two-body problem exhibits no anomalies if the 
active and passive gravitational masses are 
not equal, and that the occasionally expressed 
opinions about supposedly missing conservation 
laws are unfounded.

In the following chapter, we turn to the 
case in which, in addition to inertial and 
gravitational forces, other (binding) forces 
act. We will see that in this situation 
there can indeed be losses of conservation 
laws because the symmetries of the equations 
of motion cease to be of Noether type. 

We end this section by pointing out 
that we could have easily generalized 
equations \eqref{eq:NewGrav} to $n>2$ particles and rewritten them in 
the form \eqref{eq:NewGravAlt} after a 
redefinition of the masses as in 
\eqref{eq:DefMuMasses}. In other words, 
all arguments presented here immediately 
generalize to any finite number of particles. 
    
\section{The self-gravitating handle:
Implementing the constraints
\label{sec:SGH1}
}
Suppose the two point masses 
considered above obey the holonomic 
constraint\footnote{A constraint on the 
possible states that a system can 
attain is called holonomic (or integrable), 
if it can be expressed as 
constraints on the positions alone (rather 
than positions and velocities).} of 
being separated at constant distance. 
In pseudo-physical terms we can imagine 
that the two masses are kept apart a 
distance $d$ by a massless rod that is 
able to support any stress necessary 
to keep them at constant distance against 
their gravitational attraction. This handle 
can be used to model the bi-polar structure 
of an inhomogeneous rigid body whose parts 
contain different materials of different 
active to passive mass ratios. 
An example would be the Moon, whose 
opposite sides are rich in aluminium 
and iron respectively. In fact, accurate 
observations of the Moon's orbit by Lunar 
Laser Ranging have been used to set 
upper limits to the possible variations 
of the active to passive mass ratio for the 
respective materials. The quoted bounds are 
$4\times 10^{-12}$ in 1986 and $3.9\times 10^{-14}$ 
in 2023.\cite{Bartlett:1986,Singh.EtAl:2023}

Now, analytically, the constraint of 
constant distance $d$ is implemented 
by a constraint function $F$ whose zeros 
define the allowed configurations; that is, 
\begin{equation}
\label{eq:ConstraintFunct}
F=\vert\vec x_2-\vec x_1\vert-d=r-d\,.
\end{equation}  

Corresponding to the two forms \eqref{eq:NewGrav} and 
\eqref{eq:NewGravAlt} for the equations 
of motion, there are two ways to take 
this constraint into account. Each 
method proceeds in the standard fashion 
using Lagrange multipliers. More 
precisely, we add a term proportional 
to the $\vec x_1$-gradient of $F$ to 
the first equation (involving 
$\ddot{\vec x}_1$) and a term proportional
to the $\vec x_2$-gradient of $F$ to the 
second equation (involving 
$\ddot{\vec x}_2$). For \eqref{eq:NewGrav} 
this results in Lagrange's-Equations 
of first kind\footnote{%
I follow the traditional terminology, according 
to which the Lagrange-Equations of first kind 
are those which result directly from Newton's 
second law and the implementation of 
constraints according to the Principle of 
d'Alembert; see, e.g., \S\,12 of 
Ref.\citenum{Sommerfeld-Mechanics-2Ed}.
No assumption is made concerning the existence 
of a Lagrange function, whose Euler-Lagrange 
derivative with a $\lambda$-term added 
would be equivalent to these equations. 
If such a Lagrange function exists, then, 
in traditional terminology, the Euler-Lagrange
equations without a $\lambda$-term are said 
to be of second kind, and those with a 
$\lambda$-term are referred to as of mixed type; 
compare \S\,34 of 
Ref.\citenum{Sommerfeld-Mechanics-2Ed}.
In contrast, modern terminology does not 
seem to recognize the distinction---which is 
crucial for us---between cases with and 
without a Lagrange function. Rather, it 
always seems to assume its existence and 
simply calls  the corresponding 
Euler-Lagrange equations with a $\lambda$-term 
the Lagrange equations of the first kind, 
and those without a $\lambda$-term of the 
second kind. See, e.g., 
\url{https://en.wikipedia.org/wiki/Lagrangian_mechanics}.}
\begin{subequations}
\label{eq:NewGravConstr}
\begin{alignat}{1}
\label{eq:NewGravConstr-a}
\m{1}{i}\ddot{\vec x}_1
&\,=\,G\,\m{1}{p}\m{2}{a}
\frac{{\vec x}_2-{\vec x}_1}{\vert{\vec x}_1-{\vec x}_2\vert^3}
-\lambda\,\frac{\vec x_2-\vec x_1}{\vert\vec x_1-\vec x_2\vert}
\,,\\
\label{eq:NewGravConstr-b}
\m{2}{i}\ddot{\vec x}_2
&\,=\,G\,\m{2}{p}\m{1}{a}
\frac{{\vec x}_1-{\vec x}_2}{\vert{\vec x}_1-{\vec x}_2\vert^3}
-\lambda\,\frac{\vec x_1-\vec x_2}{\vert\vec x_1-\vec x_2\vert}
\,.
\end{alignat}
\end{subequations} 
Doing the same for \eqref{eq:NewGravAlt}
gives
\begin{subequations}
\label{eq:NewGravConstrAlt}
\begin{alignat}{1}
\label{eq:NewGravConstrAlt-a}
\mu_1\ddot{\vec x}_1
&\,=\,G\,\m{1}{a}\m{2}{a}
\frac{{\vec x}_2-{\vec x}_1}{\vert{\vec x}_1-{\vec x}_2\vert^3}
-\lambda'\,\frac{\vec x_2-\vec x_1}{\vert\vec x_1-\vec x_2\vert}
\,,\\
\label{eq:NewGravConstrAlt-b}
\mu_2\ddot{\vec x}_2
&\,=\,G\,\m{1}{a}\m{2}{a}
\frac{{\vec x}_1-{\vec x}_2}{\vert{\vec x}_1-{\vec x}_2\vert^3}
-\lambda'\,\frac{\vec x_1-\vec x_2}{\vert\vec x_1-\vec x_2\vert}
\,.
\end{alignat}
\end{subequations}  

The interesting point is that, whereas \eqref{eq:NewGrav} and \eqref{eq:NewGravAlt}
are analytically entirely equivalent, \eqref{eq:NewGravConstr} and \eqref{eq:NewGravConstrAlt} are not. 
It is for this reason that we also gave the 
Lagrange multipliers a slightly different name
($\lambda$ in the first,  $\lambda'$ in 
the second case). 

Just as the equations \eqref{eq:NewGravAlt} 
are the Euler-Lagrange equations for the 
Lagrangian \eqref{eq:LagrangeFunction},
so are the equations \eqref{eq:NewGravConstrAlt}
the Euler-Lagrange equations (of second kind) 
for the Lagrange function 
\begin{equation}
\label{eq:LagrangeFunctionConstr}
L=\frac{1}{2}\bigl(
\mu_1\vert\dot{\vec x}_1\vert^2+\mu_2\vert\dot{\vec x}_2\vert^2\bigr)
+G\,\frac{\m{1}{a}\m{2}{a}}{\vert{\vec x}_1-\vec{x}_2\vert}+\lambda'\bigl(\vert\vec x_2-\vec x_1\vert-d\bigr)\,.
\end{equation} 
Again, an alternative way to write this is 
to use \eqref{eq:PositionsExpressed} to
eliminate $\vec x_1$ and $\vec x_2$ in favor 
of ${\Vec X}_\mu$ and $\vec r$:
\begin{equation}
\label{eq:LagrangeFunctionConstr-2}
L=\frac{1}{2}\bigl(\mu_{\rm tot}\vert
{\dot{\vec X}}_\mu\vert^2+\mu_{\rm red}\vert
\dot{\vec r}\vert^2\bigr)
+G\,\frac{\m{1}{a}\m{2}{a}}{r}+\lambda'(r-d)\,.
\end{equation}
Below we will see that there is no 
Lagrangian formulation for 
\eqref{eq:NewGravConstr}. 

We shall now discuss and compare the two 
methods in more detail, starting with the 
second. From
\eqref{eq:LagrangeFunctionConstr-2} we 
observe that the additional term 
proportional to $\lambda'$ does not 
affect the invariance-properties 
of $L$ under inhomogeneous Galilean 
transformations, all of which are still 
acting by Noether symmetries. 
Hence all the ten quantities listed in 
\eqref{eq:ConsQuant} are still conserved, 
with the only difference being that the 
expression for $E$ now has an additional 
term $-\lambda'(r-d)$ on the right-hand 
side of \eqref{eq:ConsQuant-a}. The 
equations of motion are obtained by 
varying \eqref{eq:LagrangeFunctionConstr-2} 
with respect to ${\vec X}_\mu$, $\vec r$, and 
$\lambda'$. Clearly, variation with 
respect to ${\vec X}_\mu$  results again in 
\eqref{eq:EL-Eq-Alt-a}. Variation with 
respect to $\vec r$ gives an additional 
term so that \eqref{eq:EL-Eq-Alt-b} is replaced with 
\begin{equation}
\label{eq:EulerLagrangeEqConstr-1}
\ddot{\vec r}
=-G\frac{\m{1}{a}\m{2}{a}}{\mu_{\rm red}}\frac{\vec r}{r^3}+\frac{\lambda'}{\mu_{\rm red}}\,\frac{\vec r}{r}\,.
\end{equation}
Finally, variation with respect to 
$\lambda'$ gives the constraint
\begin{equation}
\label{eq:EulerLagrangeEqConstr-2}
r=d\,.
\end{equation}
Since the modulus $r$ of $\vec r$ is 
constant, $\dot{\vec r}$ must be 
perpendicular to $\vec r$. Hence there 
exists an $\vec\omega$, which, at this stage, could be time dependent, such that 
\begin{equation}
\label{eq:ConstrVel-1}
\dot{\vec r}=\vec\omega\times\vec r\,.
\end{equation}
Hence 
\begin{equation}
\label{eq:ConstrVel-2}
\ddot{\vec r}=\dot{\vec\omega}\times\vec r+\vec\omega\times(\vec\omega\times\vec r)\,.
\end{equation}
But since, according to \eqref{eq:EulerLagrangeEqConstr-1}, 
$\ddot{\vec r}\propto\vec r$, we have 
$\dot{\vec\omega}=\vec 0$ and 
$\vec\omega\cdot\vec r=0$. Hence $\vec r$
stays in the plane perpendicular to the 
constant vector $\vec\omega$ whose modulus, 
$\omega:=\vert\vec\omega\vert$, is the 
constant angular velocity with which the 
vector $\vec r$ of constant length $d$ 
rotates in that plane. In particular, 
\begin{equation}
\label{eq:RelativeAcc}
\ddot{\vec r}=-\omega^2\vec r\,.
\end{equation} 
Inserting this and 
\eqref{eq:EulerLagrangeEqConstr-2}
into \eqref{eq:EulerLagrangeEqConstr-1} 
allows the computation of the Lagrange 
multiplier:
\begin{equation}
\label{eq:LambdaSol}
\lambda'= \mu_{\rm red}\left\{
G\frac{\m{1}{a}\m{2}{a}}{\mu_{\rm red} d^2}-d\omega^2\right\} \,.
\end{equation} 
Its interpretation is that of the force 
needed to uphold the constraint. Equation \eqref{eq:LambdaSol} tells us that this
force just balances the sum of the inward pointing gravitational attraction and the 
outward pointing centrifugal force. 
Together with \eqref{eq:EL-Eq-Alt-a}, i.e.,
\begin{equation}
\label{eq:ddotXOnceMore}
{\ddot{\vec X}}_\mu=\vec 0\,,
\end{equation}
this provides all the analytic information 
that we can deduce from the equations of motion.  

Note that whereas the ${\vec X}_\mu$ still is unaccelerated, the acceleration of 
$\Vec X_i$ is given by twice differentiating 
\eqref{eq:CoM-Relation} and using  \eqref{eq:RelativeAcc} and 
\eqref{eq:ddotXOnceMore}. The result is 
\begin{equation}
\label{eq:CentreInertiaMotion}
\ddot{\vec X}_i=-\omega^2d\,S_{12}\frac{\m{\rm red}{i}}{\mu_{\rm tot}}
\,\vec n\,.
\end{equation} 
Here $\vec n$ is again the unit vector pointing from $\vec x_1$ to $\vec x_2$, as in \eqref{eq:NormalVector}. $\vec n$ rotates 
in the plane orthogonal to $\vec\omega$ with 
constant angular frequency $\omega$. It is 
only due to that rotation and the offset of 
$\vec X_i$ from ${\vec X}_\mu$ that we get 
a non-zero acceleration of $\vec X_i$.

Next we discuss the first method of 
implementing the constraint, i.e. equations 
\eqref{eq:NewGravConstr}. Multiplying 
\eqref{eq:NewGravConstr-a} by
$\m{1}{a}/\m{1}{p}$ and \eqref{eq:NewGravConstr-b} 
by $\m{2}{a}/\m{2}{p}$ gives
\begin{subequations}
\label{eq:NewGravConstrAlt2}
\begin{alignat}{1}
\label{eq:NewGravConstrAlt2-a}
\mu_1\ddot{\vec x}_1
&\,=\,G\,\m{1}{a}\m{2}{a}
\frac{{\vec x}_2-{\vec x}_1}{\vert{\vec x}_1-{\vec x}_2\vert^3}
-\lambda\,\frac{\m{1}{a}}{\m{1}{p}}\,\frac{\vec x_2-\vec x_1}{\vert\vec x_1-\vec x_2\vert}
\,,\\
\label{eq:NewGravConstrAlt2-b}
\mu_2\ddot{\vec x}_2
&\,=\,G\,\m{1}{a}\m{2}{a}
\frac{{\vec x}_1-{\vec x}_2}{\vert{\vec x}_1-{\vec x}_2\vert^3}
-\lambda\,\frac{\m{2}{a}}{\m{2}{p}}\,\frac{\vec x_1-\vec x_2}{\vert\vec x_1-\vec x_2\vert}
\,.
\end{alignat}
\end{subequations} 
We observe that now the terms proportional 
to $\lambda$ are not equal in magnitude,
in contrast to \eqref{eq:NewGravConstrAlt},
unless $S_{12}=0$. Hence, adding \eqref{eq:NewGravConstrAlt2-a} to \eqref{eq:NewGravConstrAlt2-b} and dividing by 
$\mu_{\rm tot}$ gives 
\begin{equation}
\label{eq:Constr-2MuCoM-Motion}
{\ddot{\vec X}}_\mu=-\lambda\frac{S_{12}}{\mu_{\rm tot}}
\,\vec n\,.
\end{equation}
Dividing \eqref{eq:NewGravConstrAlt2-a} 
by $\mu_1$ and 
\eqref{eq:NewGravConstrAlt2-b} by 
$\mu_2$ and subtracting the first from 
the second gives 
\begin{equation}
\label{eq:Constr-2Diff-Motion}
\ddot{\vec r}=-G\frac{\m{1}{a}\m{2}{a}}{\mu_{\rm red}}
\frac{{\vec x}_2-{\vec x}_1}{\vert{\vec x}_1-{\vec x}_2\vert^3}
+\frac{\lambda}{\m{\rm red}{i}}\,
\frac{{\vec x}_2-{\vec x}_1}{\vert{\vec x}_1-{\vec x}_2\vert}\,.
\end{equation}
As before, the constraint $\vert\vec x_2-
\vec x_1\vert=d$ leads to 
\eqref{eq:ConstrVel-1} and 
\eqref{eq:ConstrVel-2}. Since 
\eqref{eq:Constr-2Diff-Motion} still 
implies $\ddot{\vec r}\propto \vec r$ 
we again get \eqref{eq:RelativeAcc}. 
Inserting this into 
\eqref{eq:Constr-2Diff-Motion} allows 
the computation of $\lambda$:
\begin{equation}
\label{eq:LambdaSol-2}
\lambda= \m{\rm red}{i}\left\{
G\frac{\m{1}{a}\m{2}{a}}{\mu_{\rm red} d^2}-d\omega^2\right\} \,.
\end{equation} 
This is almost the same expression as 
\eqref{eq:LambdaSol}, the only difference 
being that the overall proportionality with 
$\mu_{\rm red}$ is now replaced 
with $\m{\rm red}{i}$. 

Using \eqref{eq:LambdaSol-2} in 
\eqref{eq:Constr-2MuCoM-Motion}
and also \eqref{eq:RelativeAcc}, the 
twice differentiated 
\eqref{eq:CoM-Relation} gives 
\begin{equation}
\label{eq:Constr-InertialCoM-Motion}
\ddot{\vec X}_i=-S_{12}\, G \frac{\m{1}{p}\m{2}{p}}{\m{\rm tot}{i}d^2}\,\vec n\,,
\end{equation}
which is just \eqref{eq:AccInertialCoM}, or \eqref{eq:MotionInertialCoM}, for 
$\vec r=d\vec n$. Note that the effect 
of the $\ddot{\vec r}$--term in the twice 
differentiated \eqref{eq:CoM-Relation}
is just to cancel the term $\propto\omega^2$ 
that enters in 
\eqref{eq:Constr-2MuCoM-Motion} through
\eqref{eq:LambdaSol-2}. In fact, the result  
\eqref{eq:Constr-InertialCoM-Motion} just 
equals the acceleration calculated 
from the sum of forces \eqref{eq:SumGravForces}
if one sets the distance of the mass points 
equal to $d$ and divides that by the total 
inertial mass.

The essential difference to the previous 
case is that now the $\mu$-center-of-mass 
${\vec X}_\mu$ is accelerated according to 
\eqref{eq:Constr-2MuCoM-Motion} if 
$S_{12}\ne 0$. Moreover, unlike for  \eqref{eq:NewGravConstrAlt}, there exists 
no Lagrange function $L$ so that 
\eqref{eq:NewGravConstrAlt2} are its 
corresponding Euler-Lagrange equations. Indeed, since without the $\lambda$-terms \eqref{eq:NewGravConstrAlt2} is already of 
Euler-Lagrange form, it remains to be of  
Euler-Lagrange form if and only if the 
$\lambda$-terms have that form. 
But since they only depend on positions 
and not on velocities, this is the case 
if and only if the terms proportional to 
$\lambda$ on the right-hand sides of 
\eqref{eq:NewGravConstrAlt2-a} 
and \eqref{eq:NewGravConstrAlt2-b} are the 
$\vec x_1$-gradient and  $\vec x_2$-
gradient,
respectively, of some scalar function 
$\Phi$. If such a $\Phi$ existed, adding it to \eqref{eq:LagrangeFunction} would then lead to \eqref{eq:NewGravConstrAlt2}, 
just like the $\lambda'$ terms added in \eqref{eq:LagrangeFunctionConstr} leads to 
\eqref{eq:NewGravConstrAlt}. But, in fact, it 
is easy to see  that such a function $\Phi$ 
cannot exist. Indeed, a necessary condition for its existence would be that 
\begin{equation}
\label{eq:IntegrabilityCond-1}
\frac{\partial}{\partial x_2^b}
\left(\frac{\partial\Phi}{\partial x_1^a}\right)
=
\frac{\partial}{\partial x_1^a}
\left(\frac{\partial\Phi}{\partial x_2^b}\right)\,,
\end{equation}
where 
\begin{subequations}
\begin{alignat}{1}
\label{eq:IntegrabilityCond-2a}
\frac{\partial\Phi}{\partial x_1^a}
&=-\lambda\,\frac{\m{1}{a}}{\m{1}{p}}\,
\frac{x^a_2-x^a_1}{\vert\vec x_1-\vec x_2\vert}
\,=\,\lambda\,\frac{\m{1}{a}}{\m{1}{p}}
\frac{\partial F}{\partial x_1^a}\,,\\
\label{eq:IntegrabilityCond-2b}
\frac{\partial\Phi}{\partial x_2^b}
&=-\lambda\,\frac{\m{2}{a}}{\m{2}{p}}\,\frac{x^b_1-x^b_2}{\vert\vec x_1-\vec x_2\vert}
\,=\,\lambda\,\frac{\m{2}{a}}{\m{2}{p}}
\frac{\partial F}{\partial x_2^b}\,,
\end{alignat}
\end{subequations}
with $F$ as in \eqref{eq:ConstraintFunct}.
But then \eqref{eq:IntegrabilityCond-1} is 
obviously equivalent to $S_{12}=0$. In 
other words: If $S_{12}\ne 0$ no $L$ exits 
for which \eqref{eq:NewGravConstr} 
are the Euler-Lagrange 
equations.\footnote{There exist 
general results in the literature giving
necessary and sufficient conditions for 
equations of motion to be of Euler-Lagrange 
form. Compare Theorem\,5.92 in 
Ref.\citenum{Olver:ApplicationsLieGroups}
and Sec.\,4.10, Theorem\,12, of 
Ref.\citenum{Krupka:GlobalVarGeom}.}
This does not mean that there 
are no conserved quantities. For example, from \eqref{eq:Constr-2Diff-Motion} it 
immediately follows that $\vec r\times\dot{\vec r}$ 
is constant. But, for example, the energy $E$ 
as defined in \eqref{eq:ConsQuant-a} will not be 
conserved anymore, even though equations \eqref{eq:NewGravConstrAlt2} are still 
invariant under all Galilean transformations, 
including time translations. The point is 
that they now fail to generate Noether 
symmetries.  

\section{The self-gravitating handle:
Nature of constraining force
\label{sec:SGH2}
}
In the second case we have seen that 
${\vec X}_\mu$ moves along a straight 
line according to \eqref{eq:ddotXOnceMore} 
and $\vec X_i$  rotates around 
${\vec X}_\mu$ with constant angular velocity 
$\vec\omega$ on a circle with radius 
$S_{12}\, d(\m{\rm red}{i}/\mu_{\rm tot})$.
This just means that it is the point 
${\vec X}_\mu$ rather than $\vec X_i$ on the 
line-segment between the two masses about 
which the handle rigidly rotates. The difference between these two points 
is proportional to $S_{12}$. 

In the first case, however,  ${\vec X}_\mu$ 
as well as $\vec X_i$ are accelerated. 
The acceleration of ${\vec X}_\mu$ has two contributions, one proportional to 
$\omega^2$ and the other independent of 
$\omega$. The acceleration of 
$\vec X_i$ consists just of that latter 
part and can be understood as due to 
a net total force ${\vec F}_{\rm tot}:=\vec 
F_{12}+\vec F_{21}$ acting on the total system,  which it then accelerates 
according to Newton's second law: 
\begin{equation}
\label{eq:Newton2ndLaw}
{\vec F}_{\rm tot}=\m{\rm tot}{i}\ddot{\vec X}_i\,.
\end{equation}
In particular, if initially released 
at rest without rotation, the whole 
handle will start a constantly 
accelerated motion along a straight line 
in the direction $\vec x_1-\vec x_2$ 
for $S_{12}>0$, and in the opposite 
direction $\vec x_2-\vec x_1$ for 
$S_{12}<0$. 

The explanation of this apparently 
unphysical behavior is based on 
\eqref{eq:Newton2ndLaw} and has often 
been repeated following Bondi\cite{Bondi:1957}. 
However, it may not be as obvious a 
consequence of Newtonian physics as it 
might seem at first. In fact, it uses 
the first and second axiom to calculate 
the motion caused by forces that violate
Newton's third law. And, indeed, as we 
have seen above, it rests on the tacit 
assumption that the constraint force that 
keeps the masses a fixed distance $d$ 
apart is of a particular form, namely 
that corresponding to the $\lambda$ 
terms in \eqref{eq:NewGravConstr} and 
\emph{not} that corresponding to the 
$\lambda'$ terms in \eqref{eq:NewGravConstrAlt}.
Looking at these two equations it is clear 
what the difference between these 
constraint forces is: The $\lambda$ 
terms in \eqref{eq:NewGravConstr} correspond 
to constraint forces that cause equal 
accelerations to equal inertial masses,
whereas the $\lambda'$ terms in
\eqref{eq:NewGravConstrAlt} cause equal 
accelerations to equal $\mu$-masses. 
The first case will presumably be argued 
to be the case if the binding forces 
are ``standard'' Newtonian forces, 
like a spring or, more fundamentally, 
electromagnetic forces. 

Finally, coming back to observational 
tests of the equivalence between active 
and passive gravitational mass%
\cite{Kreuzer:1968,Bartlett:1986,Singh.EtAl:2023,Erling:2023}, 
one may wonder how observationally a 
distinction is made between the ``positions'' 
${\vec X}_i$ and  ${{\vec X}}_\mu$ for 
extended objects like the Moon. In other 
words: which is the right position 
observable whose anomalous motion 
reveals $S_{12}\ne 0$, and how do we 
realize it operationally?

\section*{Mathematical Appendix}
In this appendix we collect some 
mathematical and conceptual aspects 
concerning the Galilean group and its 
implementation as group of dynamical 
symmetries of certain equations of 
motion. Our intention is to address 
readers with some elementary prior knowledge 
of group theory, who will appreciate 
more abstract lines of reasoning. 
This material is not contained in the 
journal paper and it is not essential 
for understanding the main message of 
this contribution. Rather, it 
is meant to illustrate in some detail
the meaning of the general and abstract 
concept of a dynamical symmetry group 
and how it applies to the concrete 
problem treated in the main text. 
This will enable the reader to 
1)~immediately recognize the full 
10-parameter Galilean symmetry of 
all equations of motion discussed 
in the main text, in particular 
\eqref{eq:NewGravConstr} and 
\eqref{eq:NewGravConstrAlt}, and 
2)~appreciate the fact that these 
familiar spacetime symmetries 
need not be of Noether type, i.e. 
will not always imply conserved 
quantities. In order to not overextend 
this appendix, we refer to the literature,
already cited in the main text, for 
discussions  of Noether's Theorem 
in mechanics.\cite{Desloge.Karch:1977,
Neuenschwander:2014,Scheck:Mechanics}

\subsection{%
Structure of the Galilean group}
We consider $\reals^4=\reals_t\times\reals^3_{\vec x}$ 
as model for spacetime, where $\reals_t$
represents the real line whose points label 
instants in time and $\reals^3_{\vec x}$ 
the Euclidean 3-space whose points label 
positions in space. The action of the 
Galilean group on spacetime is that 
generated by the following operations: 
\begin{itemize}
\item
Time translations by 
$b\in\reals_b$:
\begin{subequations}
\begin{equation}
\label{eq:Galilei-TT}
(t,\vec x)\mapsto (t+b,\vec x)\,.
\end{equation}
\item
Space translations by 
 $\vec a\in\reals^3_{\vec x}$:
\begin{equation}
\label{eq:Galilei-ST}
(t,\vec x)\mapsto (t,\vec x+\vec a)\,.
\end{equation}
\item
Boosts by $\vec v\in\reals^3_{\vec v}$:
\begin{equation}
\label{eq:Galilei-Boost}
(t,\vec x)\mapsto (t,\vec x+\vec vt)\,.
\end{equation}
\item
Space rotations by $R\in\SO$:
\begin{equation}
\label{eq:Galilei-Rotation}
(t,\vec x)\mapsto (t,R\vec x)\,.
\end{equation}
\end{subequations} 
\end{itemize}

Applying first \eqref{eq:Galilei-Rotation},
then  \eqref{eq:Galilei-Boost}, then 
\eqref{eq:Galilei-ST}, and finally
 \eqref{eq:Galilei-TT} gives a general
element $g$ of the Galilean group, which we 
simply label by the parameters just listed.
Hence we write  
$g:=(b,\vec a,\vec v,R)\in \reals_t\times\reals^3_{\vec a}\times\reals^3_{\vec v}\times\SO$. 
It acts on spacetime as follows (the 
action being denoted by a dot $\cdot$ 
between the acting group element and the spacetime point): 
\begin{equation}
\label{eq:GalileiAction-1}
g\cdot \bigl(t,\vec x\bigr)=\bigl(b,\vec a,\vec v,R\bigr)\cdot
\bigl(t,\vec x\bigr)=
\bigl(t+b\,,\,
R\vec x+\vec vt+\vec a
\bigr)\,.
\end{equation}
Doing this twice, first with 
$g=\bigl(b,\vec a,\vec v,R\bigr)$
followed by $g'=\bigl(b',\vec a',\vec v',R'\bigr)$, shows that 
\begin{equation}
\label{eq:GalileiAction-2}
g'\cdot\bigl(g\cdot(t,\vec x)\bigr)
=(g'g)\cdot(t,\vec x)\,,
\end{equation}
where 
\begin{equation}
\label{eq:MultGal}
\begin{split}
g'g
&=\bigl(b',\vec a',\vec v',R'\bigr)
 \bigl(b,\vec a,\vec v,R\bigr)\\
&=
\bigl(
b'+b\,,\,
\vec a'+R'\vec a+\vec v' b\,,\,
\vec v'+R'\vec v\,,\,
R'R
\bigr)\,.
\end{split}
\end{equation}
This represents the Galilean group 
product in the given parametrization. 
It is straightforward to check 
associativity, i.e. $(gg')g''=g(g'g'')$. 
The 
parameter values $(b,\vec a,\vec v,R)$
for the neutral element $g=e$ are 
\begin{equation}
\label{eq:GalileiNeutralElement}
e=
(0\,,\,
\vec 0\,,\,
\vec 0\,,\,
\mathbf{1})\,.
\end{equation}
The parameter values 
$(b',\vec a',\vec v',R')$ of the 
inverse of $(b,\vec a,\vec v,R)$ 
are obtained by equating the right-hand
side of \eqref{eq:MultGal} to \eqref{eq:GalileiNeutralElement} and 
solving for the primed parameters. 
This gives for $g^{-1}=(b,\vec a,\vec v,R)^{-1}=(b',\vec a',\vec v',R')$
\begin{equation}
\label{eq:GalileiInverse}
b'=-b,\;
\vec a'=-R^{-1}(\vec a-b\vec v),\;
\vec v'=-R^{-1}\vec v,\;
R'=R^{-1}\,.
\end{equation}
The fact expressed in Equation 
\eqref{eq:GalileiAction-2}, namely, that 
the successive application of the two 
transformations corresponding to the group 
elements $g$ and $g'$ is equal to the single 
application of the transformation associated 
with the group-product element $gg'$, is 
summarized by saying that the Galilean 
group defines an \emph{action} on 
spacetime. That notion generalizes to 
arbitrary groups and sets it may act on.

A helpful and insightful way to remember 
the above rule for group multiplication 
is to use the fact 
that the Galilean group can be embedded into 
the group $\mathrm{GL}(\reals^5)$ of real 
invertible $5\times 5$ matrices as 
follows:     
\begin{equation}
\label{eq:GalileiEmbedding}
(b,\vec a,\vec v,R)\mapsto
\begin{pmatrix}
1&0&\vec 0^\top\\
b&1&\vec 0^\top \\
\vec a&\vec v&R
\end{pmatrix}\,.
\end{equation}
Note the slight shorthand consisting in 
writing a $5\times 5$ matrix in $3\times 3$
form, with the lowest row consisting of two 
3-dimensional column vectors $\vec a$ and 
$\vec v$ and a $3\times 3$ matrix $R$. 
Matrix multiplication works exactly as usual 
and it is now easy to verify \eqref{eq:MultGal}. 
Incidentally, this also implies associativity 
for \eqref{eq:MultGal}, since matrix 
multiplication is associative. 

The trick to embed the full Galilean group 
into $\GL(\reals^5)$ may be understood as follows: 
we represent 4-dimensional spacetime, $\ST$,  
as the hyperplane in $\reals^5$ given by all 
column-vectors whose first entry is unity: 
\begin{equation}
\label{eq:SpacetimeHyperplane}
\ST=
\left\{
\begin{pmatrix}
1\\
t\\
\vec x
\end{pmatrix}:
t\in\reals,\;
\vec x\in\reals^3
\right\}\,.
\end{equation} 
Points in $\ST$ are transformed to points in 
$\ST$ under transformation of the form 
\eqref{eq:GalileiEmbedding}.
In other words, $\ST\subset\reals^5$ is an 
invariant subset under transformations 
of the form \eqref{eq:GalileiEmbedding},
which are linear in $\reals^5$ but not in
$\ST$ . In fact, $\ST$ is not a linear subspace. Vectors of the form \eqref{eq:SpacetimeHyperplane} 
cannot be added, nor is the zero vector 
contained in them. Rather, it is an 
\emph{affine} 
subspace over the linear subspace of 
vectors whose first component vanishes. 
So the embedding 
trick is useful because it allows to represent 
the affine transformations of spacetime, 
$\ST$, which is properly addressed as a 
4-dimensional \emph{affine} space, as linear 
transformations in a higher dimensional space. 

Now, what is a semidirect product? 
We will not give the general definition 
here since this would lead us more deeply 
into group-theoretic concepts that we wish 
to avoid. Rather, we concentrate on a 
specific class of examples within which the 
Galilean group falls. Let $\GL(\reals^n)$
be the group of invertible $n\times n$
matrices with real entries and 
$G\subset\GL(\reals^n)$ any subgroup. 
$G$ acts on $\reals^n$ as usual
by multiplying an $n$ component 
column-vector  $\vec v\in\reals^n$ with an 
$n\times n$ matrix $g\in G$. 
So we know what $g\vec v$ means.

Now, next to $G$ we consider another group, 
namely just $\reals^n$, which is indeed an 
abelian group if group multiplication is 
defined to be vector addition. We
consider it as the group of translations.   
The zero vector $\vec 0$ being the group's 
neutral element and the inverse of $\vec v$
being just $-\vec v$. Whenever we have two 
groups we can form the Cartesian product of 
their underlying sets; in our case 
$\reals^n\times G$. One way to make this 
into a group is to define multiplication  
component-wise, i.e. $(\vec v',g')(\vec v,g)=
(\vec v'+\vec v\,,\,g'g)$. The group 
so obtained is called the \emph{direct} 
product and also denoted by 
$\reals^n\times G$. Now, the semidirect product defines a different multiplication
rule on the same underlying set, namely
\begin{equation}
\label{eq:SEmiDirProd}
(\vec v',g')(\vec v,g)=
(\vec v'+g'\vec v\,,\,g'g)\,.
\end{equation}
Note that this differs from the direct product 
only in that $\vec v$ has an additional 
linear map $g'$ applied to it. So it is 
this additional action of $G$ on $\reals^n$
that modifies the addition of vectors in the 
first slot. The set $\reals^n\times G$ endowed 
with the multiplication rule 
\eqref{eq:SEmiDirProd} is called the 
\emph{semidirect product} of $\reals^n$ 
with $G$ and denoted by $\reals^n\rtimes G$.
It is also often called the inhomogeneous 
group $IG$, since it adds to the linear 
transformations contained in $G$ the 
translations.     

$IG=\reals^n\rtimes G$ so defined does indeed 
form a group. The neutral element is 
$(\vec 0,e)$ ($e$ being the neutral element 
in $G$) and the inverse of $(\vec v,g)$ is 
\begin{equation}
\label{eq:SEmiDirProdInv}
(\vec v,g)^{-1}=(-g^{-1}\vec v,g^{-1})\,.
\end{equation}
Finally, a straightforward calculation 
shows that \eqref{eq:SEmiDirProd} obeys 
associativity.  

Direct and semidirect products have similarities and differences in their 
group structures that should be appreciated. 
In both cases $\reals^n$ and $G$ are 
subgroups. Also, in both cases $\reals^n$
is a normal (or invariant) subgroup, 
meaning that it is  mapped into itself 
under conjugation with any element in 
the full group. Loosely speaking this 
means that $\reals^n$ has a ``unique 
position'' within the full group. 
Alternatively one could say that there 
is only one (unique) copy of $\reals^n$ 
in the group of translations within 
the full group. Now, in contrast, 
whereas $G$ is also normal in the direct 
product, that ceases to be true for the 
semidirect product.
Indeed, if we conjugate 
$(\vec 0,g)$ with $(\vec v,e)$ we get 
\begin{equation}
\label{eq:SEmiDirProdConj}
(\vec v,e)(\vec 0,g)(\vec v,e)^{-1}
=(\vec v,g)(-\vec v,e)=(\vec v-g\vec v\,,\,g)\,.
\end{equation}
Keeping $\vec v$ fixed and letting $g$ 
run through all of $G$ gives the set 
\begin{equation}
\label{eq:SEmiDirProdConjG-1}
G_{\vec v}:=
\bigl\{
(\vec v-g\vec v\,,\,g): g\in G
\bigr\}
\subset\reals^n\rtimes G\,.
\end{equation}
It is clear from the construction, but 
also easily verified directly using 
\eqref{eq:SEmiDirProd}, that this set is 
a subgroup isomorphic to $G$. Hence 
$\reals^n\rtimes G$ contains uncountably 
many \emph{different} subgroups isomorphic to $G$, 
one for each $\vec v$. They are different 
but their intersections need not be trivial:
\begin{equation}
\label{eq:SEmiDirProdConjG-2}
\begin{split}
G_{\vec v}\cap G_{\vec v'}=
\bigl\{&
(\vec v-g\vec v\,,\,g):\\
&g\in G,\ g(\vec v'-\vec v)=(\vec v'-\vec v)
\bigr\}
\end{split}
\end{equation}
This is easy to understand in the example 
of the group of Euclidean motions in 
$\reals^3$, consisting of translations
$\vec x\mapsto\vec x+\vec a$ and 
rotations $\vec x\mapsto R\vec x$. The combined 
transformation is $\vec x\mapsto R\vec x+\vec a$
and defines a semidirect product $\reals^3\rtimes\SO$
according to the multiplication rule 
$(\vec a',R')(\vec a,R)=(\vec a'+R'\vec a,R'R)$.
In that group there is a unique subgroup 
of translations, but no unique subgroup of 
rotations. Rather, there is a $\reals^3$ worth of 
such rotational subgroups, one for each point 
$\vec c$ about which the rotations take place 
(the ``center'' of rotations, which is 
always fixed). The two rotational subgroups 
fixing $\vec c$ and $\vec c'$, respectively, 
intersect in those rotations that fix the 
difference $\vec c'-\vec c$, i.e. the one-parameter 
family of rotations about the axis defined by 
$\vec c'-\vec c$.

Coming back to the semidirect product 
$\reals^n\rtimes G$, where $G$ is any subgroup 
of $\GL(\reals^n)$, we note that we can embed 
$\reals^n\rtimes G$ in $\GL(\reals^{n+1})$
as follows: 
\begin{equation}
\label{eq:SEmiDirProdEmbed-1}
(\vec v\,,\,g)\mapsto 
\begin{pmatrix}
1 & \vec 0^\top\\
\vec v & g
\end{pmatrix}\,.
\end{equation}
This is again written in a $(1+n)$ matrix 
form with ordinary rules for matrix 
multiplications. These just reproduce  
\eqref{eq:SEmiDirProd}:
\begin{equation}
\label{eq:SEmiDirProdEmbed-2} 
\begin{pmatrix}
1 & \vec 0^\top\\
\vec v' & g'
\end{pmatrix}
\begin{pmatrix}
1 & \vec 0^\top\\
\vec v & g
\end{pmatrix}
=
\begin{pmatrix}
1 & \vec 0^\top\\
\vec v'+g'\vec v & g'g
\end{pmatrix}\,.
\end{equation}

So the kind of semidirect products
discussed here are easily recognized 
in the triangular form of the matrices, 
as on the right-hand side of  
\eqref{eq:SEmiDirProdEmbed-2}. If we now 
compare this with \eqref{eq:GalileiEmbedding}
we see that in the case of the Galilean group 
we have that structure iterated: first 
for the rotation group ($R$) and the
boosts $(\vec v)$ giving rise to the 
(homogeneous) Galilean group 
\begin{equation}
\label{eq:HomogGal}
\Gal:=\reals^3_{\vec v}\rtimes\SO\,.
\end{equation}
In the next step $\Gal$ is taken and enlarged 
by another semidirect product with spacetime 
translations in 
$\reals^4=\reals_b\times\reals^3_{\vec a}$. This 
leads to the inhomogeneous Galilean group 
\begin{equation}
\label{eq:InhomogGal}
\IGal:=\reals^4\rtimes\Gal=
\bigl(\reals_b\times\reals^3_{\vec a}\bigr)
\rtimes
\bigl(\reals^3_{\vec v}\rtimes\SO\bigr)\,.
\end{equation}
This explains the systematic structure behind 
\eqref{eq:MultGal}. However, this is not 
the only way to  write $\IGal$ as a 
semidirect product.
Another one would be 
$(\reals^3_{\vec a}\times\reals^3_{\vec v})\rtimes(\reals_b\times\SO)$, or, equivalently,
$((\reals^3_{\vec a}\times\reals^3_{\vec v})\rtimes\SO)\rtimes\reals_b$, but we 
shall not need this here.\footnote{%
These other ways to write $\IGal$ as semidirect 
products become apparent if instead of 
$g=(b,\vec a,\vec v,R)$ we write 
$g=(\vec a,\vec v,b,R)$, i.e if we exchange the order 
in which time translations and boosts (which do not 
commute) act on spacetime. 
Then equation
(\ref{eq:GalileiAction-1}) 
is replaced by 
$
g\cdot (t,\vec x)
=(\vec a,\vec v,b,R)
\cdot(t,\vec x)=
(t+b\,,\,
R\vec x+\vec v(t+b)+\vec a
)
$
and (\ref{eq:MultGal})
by 
$
(\vec a',\vec v',b',R')
(\vec a, \vec v, b, R)=
(\vec a'', \vec v'', b'', R'')
$
with 
$\vec a''=\vec a'+R'\vec a-b'R'\vec v$, 
$\vec v''=\vec v'+R'\vec v$, 
$b''=b'+b$, 
and $R''=R'R$. 
This corresponds to a semidirect product 
$(\reals^3_{\vec a}\times\reals^3_{\vec v})
\rtimes(\reals_b\times\SO)$, 
where the action of 
$(b',R')\in(\reals_b\times\SO)$
on $(\vec a,\vec v)\in\reals^3_{\vec a}
\times\reals^3_{\vec v}\cong\reals^6$ 
is by $(b',R')\cdot (\vec a,\vec v)
=(R'\vec a-b'R'\vec v\,,\,R'\vec v)$. Since 
$\reals_b$ and $\SO$ commute, this 
can also be written as an iterated semidirect product
$
((\reals^3_{\vec a}\times\reals^3_{\vec v})\rtimes\SO)%
\rtimes\reals_b
$, as stated in the text. 
 }

\section{Symmetries of dynamical laws}
The general statement of a dynamical 
symmetry needs the following 
input\footnote{The terminology of 
``kinematically'' versus ``dynamically 
possible trajectories'' is borrowed from 
section 4-1 of Anderson's 
book.\cite{Anderson:PORP}}:
\begin{enumerate}
\item 
A set $\Kin$ of kinematically possible trajectories. 
\item 
A subset $\Dyn\subset\Kin$ of dynamically possible 
trajectories.
\item
A group $G$ that acts effectively\footnote{
The action of a group $G$ on a set $S$ 
is called \emph{effective}, if no element 
in $G$ but the neutral one fixes all 
points of $S$.  The requirement of acting 
effectively implies no loss of generality,
since, if it did not act effectively, and 
if $G'$ denotes the subset of elements 
in $G$ that fix all points of $S$, $G'$
is easily seen to form a normal subgroup
of $G$. Hence the quotient group $G/G'$ 
now acts effectively on $S$.} on $\Kin$.
\end{enumerate}
\emph{\noindent
The statement that a group $G$ acts as dynamical symmetries then just means 
that $G$ maps $\Dyn$ to itself; that is, 
it maps solutions to solutions.}

In the context of Newtonian mechanics, for $N$ 
point-particles, the equations are for $N$ 
$\reals^3$-valued trajectories $\vec x_a(t)$, 
$a=1,\cdots,N$, or, what amounts to the same 
statement, for a single $\reals^{3N}$-valued 
trajectory $X(t):=\bigl(\vec x_1(t),\cdots,\vec x_N(t)\bigr)$.
The equation will contain second time derivatives 
and a potential $V(X)$ which might be defined only 
in a subset $\Omega\subset\reals^{3N}$. For example, 
in case of $N$ gravitationally interacting particles,
the potential $V(\vec x_1,\cdots,\vec x_N)$ consists 
of $\frac{1}{2}N(N-1)$ terms proportional to 
the inverse distances $\vert\vec x_a-\vec x_b\vert$ 
for $a<b$. Hence the potential becomes singular 
whenever two particle positions coincide. The set 
of such points in $\reals^{3N}$ is called the 
generalized diagonal in $\reals^{3N}$ and should be 
removed. Hence the domain 
$\Omega\subset\reals^{3N}$
is given by 
\begin{equation}
\label{eq:DomainV}
\Omega:=\bigl\{(\vec x_1,\cdots,\vec x_N)\in\reals^{3N}:
\vec x_a\ne\vec x_b\,\forall a\ne b\bigr\}\,.
\end{equation}
The group $\IGal$ acts on the time axis 
$\reals_t$ (just by time translation) and also 
on space $\reals^3_{\vec x}$, albeit on the latter 
in a time-dependent  fashion (because of the boosts). 
This implies that $\IGal$ also acts on the set 
of maps from $\reals_t$ to 
$\reals^3_{\vec x}$. This is because maps (or 
functions) are in one-to-one correspondence with graphs, 
which are subsets of the Cartesian product space 
$\reals_t\times\reals^3_{\vec x}=\reals^4$. 
But since $\IGal$ acts on that $\reals^4$ in 
the way discussed above, which has the property 
that it maps a graph again to a graph 
(as one easily sees), it also acts on the 
set of mappings. Moreover, as seen above, 
each group element just shifts or rotates
within $\reals^3_{\vec x}$, which means 
that each group element acts by a 
bijection in $\reals^3_{\vec x}$,
so that two non-coinciding points 
are always mapped to non-coinciding 
points. Finally, since that action
is smooth, there is no loss of 
differentiability for any curve 
$\vec x(t)$ after being mapped by an 
element in $\IGal$. Taken together, all
this implies that $\IGal$ acts on the 
space $C^2(\reals,\Omega)$ of twice 
continuously differentiable curves 
in $\Omega$. That function space
is our set of kinematically possible 
trajectories: 
\begin{equation}
\label{eq:DefKPT}
\Kin:=C^2(\reals,\Omega)\,.
\end{equation} 
From what has been said above the action can actually 
be deduced. Let 
$\vec x_a:\reals_t\rightarrow\reals_{\vec x}$ be the 
trajectory of any of the particles, and 
$g=(b,\vec a,\vec v,R)\in\IGal$. Then the image
$g\cdot\vec x_a$ of this trajectory under $g$ is 
given by 
\begin{equation}
\label{eq:ActionGalPath-1}
(g\cdot\vec x_a)(t)=
R\,\vec x_a(t-b)+\vec v(t-b)+\vec a\,.
\end{equation}   
Applying this rule twice shows by direct 
calculation that this is indeed an action 
of $\IGal$: 
\begin{equation}
\label{eq:ActionGalPath-2}
g'\cdot(g\cdot\vec x_a)=(g'g)\cdot\vec x_a\,.
\end{equation}  

This equation expresses the fact that 
\eqref{eq:ActionGalPath-1} actually 
defines an action of the group $\IGal$ on 
the set of kinematically possible 
one-particle
trajectories. Note that it is not sufficient 
to implement each group element as a 
transformation in some ad hoc way; rather, 
these implementations must fit together to 
form an action of the entire group. Since 
this 
point is not trivial and often not given 
sufficient attention, we shall here display 
the simple proof. We set 
$\vec x'_a:=g\cdot\vec x_a$ and $\vec x''_a:=g'\cdot\vec x'_a$. A first 
application of \eqref{eq:ActionGalPath-1}
then gives
\begin{equation}
\label{eq:ActionGalPath-3}
(g'\cdot\vec x'_a)(t)=
R'\,\vec x'_a(t-b')+\vec v'(t-b')+\vec a'\,.
\end{equation}   
In this expression we replace 
$\vec x'_a(t-b')$ by a second application of 
formula \eqref{eq:ActionGalPath-1}, 
in which we replace the argument 
$t$ by $t-b$. This leads to 
\begin{equation}
\label{eq:ActionGalPath-4}
\begin{split}
\vec x''_a(t)
=\,&R'\,[R\vec x_a(t-b'-b)+\vec v(t-b'-b)+\vec a]\\
&+\vec v'(t-b')+\vec a'\\
=\,&R''\,\vec x_a(t-b'')+\vec v''(t-b'')
+\vec a''\,,
\end{split}
\end{equation}   
with $b''=b'+b$, $\vec a''=\vec a'+R'\vec a+\vec v'b$, $\vec v''=\vec v'+R'\vec v$, and 
$R''=R'R$, just as in \eqref{eq:MultGal}. Note in particular the origin of the term $\vec v'b$ in the expression of $\vec a''$
and that in \eqref{eq:ActionGalPath-1}  
it is essential to have $(t-b)$ in the argument of $\vec x_a$ (so as to shift 
that function in the \emph{positive} 
time direction for $b>0$) and also in the 
term $(t-b)$ multiplying $\vec v$.  

Now, finally, this action trivially 
extends to the $N$-fold Cartesian product 
$X(t)=\bigl(\vec x_1(t),\cdots,\vec x_N(t)\bigr)$ of one-particle curves and from 
there restricts to 
$\Kin=C^2(\reals,\Omega)$.
Given the $N$-body gravitational problem of the main 
text (generalized from $N=2$), it is obvious and needs 
no calculation that the action 
\eqref{eq:ActionGalPath-1} of $\IGal$ 
to each of the $N$ trajectories indeed implements 
$\IGal$ as symmetry group of the 
combined set of equations.  

There may be an issue with the implementation 
of time translations in case we admit 
solutions that run out of $\Omega$ in finite time.
This happens if pairs of particles collide in 
finite time. In this case there will be no global 
time-translation invariance,  i.e., for all values 
$b\in\reals_b$, as would be required if 
$\reals_b$ acted as a group.


\end{document}